\documentclass[twocolumn,english,aps,prd,prl,longbibliography]{revtex4-2}
\pdfoutput=1

\usepackage[T1]{fontenc}
\usepackage[a4paper]{geometry}
\usepackage{hyperref}
\usepackage{babel}
\usepackage{amsmath}
\usepackage{tikz}\usepackage{tikz-feynman}


\newcommand{\eq}[1]{\begin{equation}#1\end{equation}}
\newcommand{\eqs}[1]{\begin{equation}\begin{split}#1\end{split}\end{equation}}

\begin{document}

\title{On-shell Bootstrap for n-gluons and gravitons scattering in (A)dS,\\ Unitarity and Soft limits}

\author{Jiajie Mei$^{a}$}
\author{Yuyu Mo$^{b}$}
\affiliation{$^{a}$ Department of Mathematical Sciences, Durham University, \\
Stockton Road, DH1 3LE Durham, UK}

\affiliation{$^{b}$ Higgs Centre for Theoretical Physics, School of Physics and Astronomy,\\ The University of Edinburgh, Edinburgh EH9 3FD, Scotland, UK}

\begin{abstract}
We propose an algorithm to recursively bootstrap $n$-point gluon and graviton Mellin-Momentum amplitudes in (A)dS spacetime using only three-point amplitude. We discover that gluon amplitudes are simply determined by factorization for $n\geq 5$. The same principle applies to $n$-point graviton amplitudes, but additional constraints such as flat space and soft limits are needed to fix contact terms. Furthermore, we establish a mapping from $n$-point Mellin-Momentum amplitudes to $n$-point cosmological correlators. We efficiently compute explicit examples up to five points. This leads to the first five-graviton amplitude in $AdS_{d+1}$.
\end{abstract}

\maketitle

\section{Introduction}
Einstein Gravity is notoriously challenging to compute, even at the perturbation level. Conversely, the modern scattering amplitudes approach in flat space has achieved tremendous success by employing the on-shell approach to compute Gravity amplitudes. More generally, the tree-level S-matrix is entirely constrained by Lorentz invariance, locality, unitarity, and gauge redundancy\cite{Cheung:2017pzi,Benincasa:2007xk}. Moreover, the BCFW recursion \cite{Britto:2005fq} significantly enhances accessibility to higher-point tree-level Gravity amplitudes, requiring only three-point amplitude as input. This eliminates the need to know the infinite expansion of the Einstein-Hilbert action.\\
However, such comprehension in curved space is still in its early stages, yet undeniably crucial for understanding our Universe—especially the uniqueness of Einstein Gravity in curved space. As a small step towards unravelling Gravity in curved space, we explore this bootstrap approach in the following maximally symmetric spacetime, (Anti)de-Sitter space.
While significant progress has been made in the cosmological bootstrap program\cite{Baumann:2022jpr,Arkani-Hamed:2018kmz,Baumann:2020dch,Pajer:2020wxk,Goodhew:2020hob,Melville:2021lst,Jazayeri:2021fvk,Albayrak:2019yve,Albayrak:2020fyp}, exploration of spinning particles and amplitudes beyond four-point remains very limited\cite{Alday:2023kfm,Cao:2023cwa,Chu:2023kpe,Bissi:2022wuh,Li:2023azu,Li:2022tby,Armstrong:2020woi,Goncalves:2019znr,Chen:2023xlt,Goncalves:2023oyx}. Building on insights from flat space, it becomes evident that grasping the structure of amplitudes in curved space requires a deep understanding of higher-point amplitudes.

In this letter, we introduce a novel and efficient algorithm for bootstrapping $n$-point amplitudes, incorporating the modern on-shell amplitude approach. The key concept involves recycling lower-point on-shell amplitudes to recursively construct higher-point amplitudes. Then, by taking the residue of OPE poles, we fix the amplitude up to contact terms. Finally, by comparing with the soft and flat space limit, we determine all the contact terms. It is crucial to emphasize that our algorithm is entirely automated and requires no guesswork, enabling possible the exploration of unknown higher-point function. Our method proves valuable not only for comprehending the structure of higher-point Mellin-Momentum amplitudes, but can also easily map the result from amplitude to Cosmological correlator.\\
The paper is organized as follows: In Section \ref{section2}, we review the Mellin-Momentum space formalism. In section \ref{section3}, we write down the general structure of $n$-point Mellin-Momentum amplitude, and we show that how we can recursively bootstrap out the amplitude according to their pole structures. Then as a warm-up of our method, in section \ref{section4} and \ref{section5} we recursively bootstrap out the four and five point gluon amplitudes and four graviton amplitude. In section \ref{section6} we derive the first five graviton amplitude. In section \ref{section7}, we give a clear mapping to map $n$-point Mellin-Momentum amplitude to correlators in momentum space.
\section{Mellin-Momentum amplitude}\label{section2}
We are considering Yang-Mills theory and Einstein Gravity in $AdS_{d+1}$, with the convention of metric:
\begin{equation}
\tilde{g}_{mn}dx^{m}dx^{n}=\tfrac{\mathcal{R}^{2}}{z^{2}}(dz^{2}+\eta_{\mu\nu}dx^{\mu}dx^{\nu}),
\end{equation}
with $0 <z<\infty$ and $\mathcal{R}=1$. The spinning particles can be parameterized as a scalar-like solution. For Yang-Mills, we set $\mathbf{A}_{m}=(\mathcal{R}/z) A_{m}$. For spin-2, the graviton will be parametrized as $g_{mn}=\tilde{g}_{mn}+\frac{\mathcal{R}^2}{z^2}h_{mn}$. Then the free equations of motion are given by:
\eqs{\mathcal{D}_k^{d-1} A_\mu(k, z)&=0,\\
\mathcal{D}_k^d h_{\mu \nu}(k, z)&=0, \label{freeEoM}
}
where the equation of motion (EoM) operator in momentum space is defined as:
\begin{align}\label{eq:eom}
    &\mathcal{D}_{k}^{\Delta} \phi_{\Delta} (k,z) =0 ,\nonumber \\
    &\mathcal{D}_{k}^{\Delta}  \equiv   z^{2}k^{2}-z^{2}\partial_{z}^{2}-(1-d)z\partial_{z}+\Delta(\Delta-d),
\end{align}
with  $\Delta$ the scaling dimension and $k=|\vec{k}|$ the norm of the boundary momentum.
So spinning particles are just scalar with polarization dressed-up: $A_{\mu}(k,z)=\varepsilon_{\mu}\phi_{d-1}(k,z)$ and $h_{\mu \nu}(k,z)=\varepsilon_{\mu}\varepsilon_{\nu}\phi_{d}(k,z)$. Now consider the Mellin-Fourier transform $\phi(x,z)\sim e^{ik\cdot x}z^{-2s+d/2}\phi(s,k)$ \cite{Sleight:2019hfp,Sleight:2019mgd,Sleight:2021iix}. Our target is to compute the Mellin-Momentum amplitude $\mathcal{A}_n(s_i,k_i,z)$ \cite{Mei:2023jkb}, which is defined as:
\begin{align}\label{eq:MMamplitude}
  \Psi_n & =  \int [ds_i]\int \frac{dz}{z^{d+1}} \mathcal{A}_n(s_i,k_i,z)\prod_{i=1}^n\phi(s_i,z),
\end{align}
with $\Psi_n=\delta^d(\Vec{k}_T) \langle \mathcal{O}(\Vec{k}_1) \dots \mathcal{O}(\Vec{k}_n) \rangle $ being the CFT boundary correlators in momentum space or the wavefunction coefficient in dS after Wick rotation. This definition of Mellin-Momentum amplitude shares the similar analytic structure of the S-matrix in flat space\footnote{See also \cite{Melville:2023kgd,Donath:2024utn} for defining S-matrix in de Sitter space, where the optical theorem is closer to S-matrix in Minkowski space.}, which makes the on-shell bootstrap to $n$-point possible. Hence, we can also derive new Feynman rules for gluons with the same structure as flat space, which are recorded in the Appendix \ref{appendixA}. 

\section{Amplitude Bootstrap}\label{section3}
We follow the same logic of amplitude bootstrap as in Minkowski space. Thus, our only input for the AdS amplitudes is the following 3-pt on-shell Yang-Mills amplitude:
\begin{equation}
   \mathcal{A}_3=z(\varepsilon_1 \cdot \varepsilon_2 \varepsilon_3 \cdot k_1+\varepsilon_2 \cdot \varepsilon_3 \varepsilon_1 \cdot k_2 + \varepsilon_3 \cdot \varepsilon_1 \varepsilon_2 \cdot k_3).\label{ymamp3}
\end{equation}
Subsequently, the $n$-point amplitudes are determined by factorization, soft limits (OPE limits), and flat space limits, based on their pole structures.\\
Mellin-Momentum amplitudes only exhibit two types of poles. By examining the pole structure of the propagator we found that the general structure of $n$-point amplitude is given as follows:
\begin{align}
    \mathcal{A}_n=& \frac{a_1(12\dots n)}{\mathcal{D}^{\Delta}_{k_I}\mathcal{D}^{\Delta}_{k_J}\dots \mathcal{D}^{\Delta}_{k_M}}+ \frac{a_2(12\dots n)}{\mathcal{D}^{\Delta}_{k_J}\dots \mathcal{D}^{\Delta}_{k_M}}+\dots \nonumber \\
    &+\frac{b(12\dots n)}{ k_I^{2m} k_J^{2m} \dots k_M^{2m}}+\dots +c(12\dots n),
\end{align}
where the order of OPE poles, for gluon $m=1$ and graviton can be $m=1,2$. Next, we will see how each of the steps fixes the amplitude by pole structure.\\
\textbf{Factorization}: Unitarity implies that the amplitude will factorize into lower point on-shell amplitudes on the factorization pole $\frac{1}{\mathcal{D}^{\Delta}_{k_I}}$ \cite{Goodhew:2020hob,Goodhew:2021oqg,Meltzer:2020qbr}\footnote{The factorization pole can be understood similar to flat space, and it occurs when the inverse two-point function blows up from Eq(\ref{freeEoM})}:
\eq{
\mathcal{A}_n \to \frac{\sum_h \mathcal{A}_{a} \mathcal{A}_{n-a+2}}{\mathcal{D}^{\Delta}_{k_I}}.
}
This step will fix all the $a_i$ terms in our ansatz.\\
\textbf{Internal Soft limit (OPE limit)}: When two operators get close to each other, it implies that the internal soft momentum $k_I^2 \to 0$, whose behaviour is then controlled by lower-point amplitudes from the usual OPE analysis\cite{Arkani-Hamed:2015bza,Sleight:2019hfp,Bzowski:2022rlz,Assassi:2012zq}, and it implies the residue of the OPE poles must be zero:
\eqs{
 \underset{k_I^2 \to 0}{\mathrm{Res}}\mathcal{A}_n = 0.
}
Remarkably, with this condition, all the $b_i$ terms in our ansatz are simply determined by taking the residue of $a_i$:
\eqs{
{b(12\dots n)}=-\underset{k_I^2 \to 0}{\mathrm{Res}}\left( \frac{a_1(12\dots n)}{\mathcal{D}^{\Delta}_{k_I}\mathcal{D}^{\Delta}_{k_J}\dots \mathcal{D}^{\Delta}_{k_M}}+\dots \right).
}
\textbf{Flat space limit}: In the flat space limit, the particles are ignoring the curvature correction and the Lorentz symmetry will emerge and give us the S-matrix in flat space\cite{Arkani-Hamed:2018bjr,Raju:2012zr,Penedones:2010ue,Li:2021snj}:
\eqs{
\lim_{z \to \infty} \mathcal{A}_n \to A_n.
}
This step fixes all the contact terms $c_i$ in our ansatz.
For $n\geq 5$ gluon amplitudes, the $n$-point functions are fully determined by the $(n-1)$ point amplitudes through factorization and the residue of the OPE poles. There are no contact terms beyond 4-point, as this would result in an incorrect flat space limit.\\
Gravity is slightly more subtle, but the same procedure applies, leaving us with only the unfixed contact interaction. Subsequently, these contact terms are fully determined by the flat space limit and external soft limit. To elaborate, we perform dimensional reduction on $n$-point graviton amplitudes, setting: $\varepsilon_i \cdot \varepsilon_j=1, \varepsilon_i \cdot k=0, \varepsilon_j \cdot k=0$ resulting in two scalar and $n-2$ graviton amplitudes. Then by taking the momentum of the scalar soft $k_i \to 0$, this amplitude will vanish due to shift symmetry\footnote{To be more precise, after dimensional reduction we obtain $\langle \phi \phi hhh\dots \rangle $ where the scalar will couple with graviton as $\nabla^m \phi \nabla^n \phi h_{mn} $, which enjoys a protected shift symmetry $\phi \to \phi +C$}, which is the Alder zero in curved space\cite{Armstrong:2022vgl,Armstrong:2022csc}. Our procedure can be fully automated to $n$-point as we will now demonstrate.
\section{Gluon amplitudes}\label{section4}
Following our amplitude bootstrap procedure, we now start with 3-point Yang-Mills amplitude in Eq \eqref{ymamp3} to bootstrap  the color ordered four-point amplitude:
\begin{equation}\label{eq:4YM}
\mathcal{A}_4=\frac{a(1,2,3,4)}{\mathcal{D}_{k_{s}}^{d-1}} + \frac{b(1,2,3,4)}{k_{s}^2}+c(1,2,3,4)+[t],
\end{equation}
where $[t]$ denotes the t-channel contribution and $k_s^2=k^2_{12}=(\Vec{k}_1+\Vec{k}_2)^2$. Then by factorization:
\begin{align}
    a(1,2,3,4)=\sum_{h=\pm} \mathcal{A}_3(1,2,-k_I^h) \mathcal{A}_3(k_{I}^{-h},3,4),
\end{align}
where the sum over of helicity is given by:
\eqs{
\sum_{h=\pm} \varepsilon_{\mu}(k,h) \varepsilon_{\nu}(k,h)^* &=\eta_{\mu \nu}-\frac{k_{\mu}k_{\nu}}{k^2} \equiv \Pi_{\mu \nu}.
}
Simply taking the residue of the OPE pole, we obtain:
\begin{align}
    b(1,2,3,4)=-\underset{k_{s}^2 \to 0}{\mathrm{Res}}\frac{a(1,2,3,4)}{\mathcal{D}_{k_{s}}^{d-1}}.
\end{align}
Finally, $c(1,2,3,4)$ is fixed by the flat space limit, which is simply the four-point contact terms. This result matches with our new Feynman rules result in Appendix [\ref{appendixA}] and literature \cite{Mei:2023jkb,Albayrak:2018tam}.\\
We now turn to the five-point ansatz:
\begin{equation}
    \mathcal{A}_5=
     \frac{a_1(1,2,3,4,5)}{\mathcal{D}_{k_{12}}^{d-1} \mathcal{D}_{k_{45}}^{d-1}}+ \frac{a_2(1,2,3,4,5)}{\mathcal{D}_{k_{12}}^{d-1}}+ \frac{b(1,2,3,4,5)}{k_{12}^2k_{45}^2}+Cyc,
\end{equation}
where $Cyc$ means sum over cyclic permutation, and factorization fixes all the $a$ terms by recycling the four-point Eq (\ref{eq:4YM}):
\eqs{
    &a_1(1,2,3,4,5)=\sum_{h} a(1,2,3,-k_I)\cdot \mathcal{A}_3(k_{I},4,5), \\ 
    &a_2(1,2,3,4,5)\\ 
    =&\sum_{h} \mathcal{A}_3(1,2,-k_I) \cdot\left(\frac{b(k_I,3,4,5)}{k_{45}^2}+c(k_I,3,4,5)+[t]\right).
}
Everything is fixed by the four-point on-shell amplitude so far. The final step is again the OPE poles, and we simply need to consider the residue on double OPE limits, which takes a very simple form:
\begin{footnotesize}
    \begin{align}
& -b(1,2,3,4,5)\notag\\ \nonumber
= & \underset{
    \begin{array}{l}
    \substack{\scriptscriptstyle 
 k^2_{12} \to 0}\\
      \substack{ \scriptscriptstyle  k^2_{45} \to 0}
    \end{array} }{\mathrm{Res}} \left (\frac{a_1(1,2,3,4,5)}{\mathcal{D}_{k_{12}}^{d-1} \mathcal{D}_{k_{45}}^{d-1}}+\frac{a_2(1,2,3,4,5)}{\mathcal{D}_{k_{12}}^{d-1}}+\frac{a_2(4,5,1,2,3)}{\mathcal{D}_{{k_{45}}}^{d-1}} \right) \notag  \\
=&\frac{\left(s_1-s_2\right)\left(s_4-s_5\right)}{z^3}\varepsilon_1\cdot \varepsilon_2 \varepsilon_4\cdot \varepsilon_5\varepsilon_3\cdot (k_4+k_5) .
\end{align}
\end{footnotesize}
We have obtained the full color-ordered five-point amplitude. It is easy to verify that the flat space limit is correct. Additionally, we have confirmed that it exactly matches the results from the new Feynman rules calculation and the literature \cite{Albayrak:2019asr,Albayrak:2018tam} after applying the mapping in Section \ref{section7}.\\

\section{Four Graviton amplitude}\label{section5}
The four-graviton correlator in $dS_4$ was first computed in \cite{Bonifacio:2022vwa}.  In this section, we will demonstrate how our bootstrap method enables efficient computation of this amplitude in $AdS_{d+1}$, beginning with the three-point amplitude:
\begin{align}
    \mathcal{M}_3=\mathcal{A}_3^2.
\end{align}
Our four-point ansatz is given as:
\begin{equation}
\mathcal{M}_4=\frac{a(1,2,3,4)}{\mathcal{D}_{k_{s}}^{d}} + \frac{b(1,2,3,4)}{k_{s}^{2m}}+c(1,2,3,4)+\mathcal{P}(2,3,4),\label{ampgr4}
\end{equation}
where remember that $m$ can be 1 or 2 and $\mathcal{P}(2,3,4)$ denotes permutation to obtain other channels. Similar to the spin-1 case, factorization and OPE limit gives:
\eqs{
    a(1,2,3,4)&=\sum_{h=\pm} \mathcal{M}_3(1,2,-k_I^h) \cdot \mathcal{M}_3(k_I^{-h},3,4) ,\\
    b(1,2,3,4)&=-\underset{k_{s}^2 \to 0}{\mathrm{Res}} \frac{a(1,2,3,4)}{\mathcal{D}_{k_{s}}^d},
}
where the spin-2 polarization sum is given by,
\eqs{
   & \sum_{h=\pm} \varepsilon_{\mu \nu}(k,h) \varepsilon_{\rho \sigma}(k,h)^* \\ 
   =&\frac{1}{2} \Pi_{\mu \rho}\Pi_{\nu \sigma}+ \frac{1}{2}\Pi_{\mu \sigma} \Pi_{\rho \nu}-\frac{1}{d-1}  \Pi_{\mu \nu}\Pi_{\rho \sigma}.
}
Now we are only left with contact terms in our bootstrap ansatz. Since Einstein Gravity is a two derivatives theory, the structure of contact terms is quite simple and can only have up to two derivatives\footnote{The most streamlined approach to deriving them, without any guesswork, would be to subtract their flat space limit and add the remaining terms into the ansatz. However, these terms will include unfixed sub-leading curvature corrections. Subsequently, the external soft limit will resolve all curvature corrections, similar to enhanced soft limits in EFT \cite{Armstrong:2022vgl,Armstrong:2022csc}}. We can classify them into two types:
\begin{equation}
c(1,2,3,4)=c_0(1,2,3,4)+c_1(1,2,3,4).
\end{equation}
For $c_0$, it originates from a part of the graviton propagator, which is necessary for the amplitude to have Lorentz invariance in flat space limit and the curvature correction can easily be determined by external soft limit,
\eqs{
    &c_0(1,2,3,4) \\
    =&\varepsilon^2_{12,34}\frac{8d(s_1-s_2)(s_3-s_4)-4(s_1+s_3-s_2-s_4)^2+d^2}{16(d-1)},\label{c0explain}
}
where we used notation $\varepsilon_{12,34}\equiv\varepsilon_1 \cdot \varepsilon_2 \varepsilon_3 \cdot \varepsilon_4$. Finally, $c_1$ is the four-point contact interaction, whose general form will be:
\eqs{
    c_1(1,2,3,4)=&\varepsilon_{ab,cd,ef}(C_1 z^2\varepsilon_m \cdot k_i \varepsilon_n \cdot k_j +C_2\varepsilon_m \cdot \varepsilon_n \mathcal{D}_{ks}^d),
}
which is simply two/zero derivatives contact interaction. We can readily determine its coefficient using the flat space limit \footnote{One might be concerned that the two-derivative contact interaction could have $1/R^2$ correction, which vanishes in the flat space limit. However, this term would not exhibit the correct behavior under external soft limits.}. This completely determines the four-graviton amplitude and matched with \cite{Mei:2023jkb}, and \cite{Bonifacio:2022vwa,Armstrong:2023phb} after using the mapping in section \ref{section7}.


\section{Five Graviton amplitude}\label{section6}
To demonstrate the power of our algorithm, we will use this method to bootstrap the first five-graviton amplitude in $AdS_{d+1}$, the five-point ansatz is given by:
\begin{align}
   & \mathcal{M}_5=
     \frac{a_1(1,2;3;4,5)}{\mathcal{D}_{k_{12}}^{d} \mathcal{D}_{k_{45}}^{d}}+ \frac{a_2(1,2;3,4,5)}{\mathcal{D}_{k_{12}}^{d}}\\ \nonumber
     +& \frac{b_1(1,2;3;4,5)}{k_{12}^{2m_1}k_{45}^{2m_2}}+\frac{b_2(1,2;3,4,5)}{k_{12}^{2m}}+c(1,2,3,4,5)+Perm.
    \end{align}
 We can recycle our four-point result to obtain all the terms with factorization poles:
\eqs{
    &a_1(1,2;3;4,5)=\sum_h a(1,2,3,-k_I)\cdot \mathcal{M}_3(k_{I},4,5), \\
    &a_2(1,2;3,4,5)\\ 
    =&\sum_{h} \mathcal{M}_3(1,2,-k_I) \cdot\left(\frac{b(k_I,3,4,5)}{k_{45}^{2m}}+c(k_I,3,4,5)+\mathcal{P}(3,4,5)\right).
}
To simplify the calculation we first fix the terms with double OPE poles, which is the same procedure as Yang-Mills\footnote{Technically, since $m$ could be 1 or 2, there are four terms with double poles for $b_1$, but since they can all be obtained by taking residue on both poles, we keep this notation for convenience.}:
    \begin{align}
&-b_1(1,2;3;4,5)\notag\\ 
=&\underset{
    \begin{array}{l}
    \substack{\scriptscriptstyle k^2_{{12}} \to 0}\\
      \substack{ \scriptscriptstyle 
 {k^2_{{45}} \to 0}}
    \end{array} }{\mathrm{Res}}\left(\frac{a_1(1,2;3;4,5)}{\mathcal{D}_{k_{{12}}}^{d} \mathcal{D}_{k_{{45}}}^{d}}+\frac{a_2(1,2;3,4,5)}{\mathcal{D}_{k_{{12}}}^{d}}+\frac{a_2(4,5;1,2,3)}{\mathcal{D}_{{k_{{45}}}}^{d}} \right ) .
\end{align}
However, unlike Yang-Mills, gravity includes terms with single OPE poles. Furthermore, as multiple channels contribute to the same single OPE poles, it becomes imperative to combine different channels. Strikingly, we can still resolve this issue simply by taking residues:
    \eqs{
&-b_2(1,2;3,4,5)\\ 
=&\underset{\substack{\scriptscriptstyle 
 k^2_{{12}} \to 0}}{\mathrm{Res}}\left\{ \frac{a_1(1,2;3;4,5)}{\mathcal{D}_{k_{{12}}}^{d} \mathcal{D}_{k_{{45}}}^{d}}+\frac{b_1(1,2;3;4,5)}{k_{12}^{2m_1} k_{45}^{2m_2}} +Cyc(3,4,5)\right.\\
&\left.+\frac{a_2(1,2;3,4,5)}{\mathcal{D}_{k_{{12}}}^{d}}+\left(\frac{a_2(3,4;5,1,2)}{\mathcal{D}_{k_{{34}}}^{d}}+Cyc(3,4,5)\right)\right \}.
}
Finally, we are left with only contact terms, which can be resolved by following the same procedure as for the four-point case,
\eqs{
c_0(1,2;3,4,5)=&\frac{d^2-4(s_1-s_2)^2-2d(s_1+s_2)}{16(d-1)} \\ & \times \varepsilon_{12,12,34,35,45}.
}
Similar to four-point, the five-point contact interaction takes the following form, and we are left only with coefficients that can be readily determined by comparison with the flat space amplitude,
\eqs{
    c_1=&\varepsilon_{ab,cd,ef,rs}(C_1 z^2\varepsilon_m \cdot k_i \varepsilon_n \cdot k_j +C_2\varepsilon_m \cdot \varepsilon_n \mathcal{D}_{k_{ij}}^d).\label{fiveptc1}
}
Summing over permutation for all the terms above, we obtained the first five-graviton amplitude in $AdS_{d+1}$ and it shares the similar analytic structure of S-matrix in flat space.

\section{Mapping to Momentum space}\label{section7}
Ultimately, our interest still lies in correlators composed of pure kinematic momentum. Mellin-Momentum amplitude not only serves as a convenient framework for understanding amplitude structure but also serves as a useful computational tool for cosmological correlators. In this section, we will verify all expressions in the paper by mapping them back to momentum space. We will provide a straightforward algebraic algorithm to demonstrate that this transition is transparent and simple for $n$-point without doing any integrals. For Yang-Mills in $d=3$, Mellin variable $s_i$ by definition gives $s_i \to \frac{z k_i}{2}+\frac{1}{4}$, then the $n$-point amplitude can be mapped as follows:
\eqs{\text{Amplitudes}& \to \text{Correlators} \\
z^{4-n}A(k,\varepsilon)& \to \frac{A(k,\varepsilon)}{E_t} \\
\frac{z^{6-n}A(k,\varepsilon)}{\mathcal{D}^{d-1}_{k_I}}& \to \frac{A(k,\varepsilon)}{E_{t}E_{I}E_{t-I}}\\
\frac{z^{n-2} A(k,\varepsilon)}{\mathcal{D}^{d-1}_{k_I} \mathcal{D}^{d-1}_{k_J} \dots \mathcal{D}^{d-1}_{k_M}}& \to \sum_{\sigma}\frac{1}{E_{t}E_{I}} \frac{A(k,\varepsilon)}{\mathcal{D}^{d-1}_{k_J} \dots \mathcal{D}^{d-1}_{k_M}},
}
where the sum over $\sigma$ represents summing over all the possible permutation on ${I,J,\dots,M}$. Our notation are total energy pole $E_t=\sum_{i=1}^{n} k_i$, and sub-total energy pole $E_I$. Such recursive integral is not too surprising for Yang-Mills, given their Weyl invariance in $3+1$ dimension\cite{Baumann:2021fxj,Arkani-Hamed:2017fdk}.
However, for Gravity in $d=3$, we found a similar recursion for $n$-point scalar integral as well! 
\eqs{
\text{Amplitudes}& \to \text{Correlator} \\    z^2 M(k,\varepsilon) & \to M(k,\varepsilon) \mathcal{C}_1 \\
\prod\limits_{m=1}^{l}\left(-2s_{m}+d/2\right) M(k,\varepsilon) & \to M(k,\varepsilon)\mathcal{C}_2^{l} \label{candp}\\ \frac{M(k,\varepsilon)}{\mathcal{D}^d_{k_I}}  & \to M(k,\varepsilon)\mathcal{I}_I \\
    \frac{M(k,\varepsilon)}{\mathcal{D}^d_{k_J}\dots \mathcal{D}^d_{k_M}} & \to M(k,\varepsilon)\mathcal{I}_{J\dots M}.
}
The first contact integral with no derivatives is:
\eqs{
\mathcal{C}_1^{(n)}= \left(\sum\limits_{m=0}^{n-2}m!\sum\limits_{1\le i_1<\ldots<i_{m+2}}^{n} \frac{k_{i_1} \ldots k_{i_{m+2}}}{E_t^{m+1}}\right)-E_t .\label{contact1}
} 
For number of derivatives greater than two ($l \geq 2$),
\eqs{
\mathcal{C}_2^{l;(n)}=\Bigg(\sum_{m=0}^{n-l} {(2l-4) !} \sum_{\begin{array}{c}
  \substack{\scriptscriptstyle l+1\le i_{l+1}<} \\  \substack{\scriptscriptstyle  \ldots<i_{m+l}}
\end{array}}^n \frac{k_{i_{l+1}} \ldots k_{i_{m+l}}}{E_t^{(2l-4)+m+1}}\Bigg)(-1)^lk_1^2...k_l^2,\label{contact2}
}where when $m=0$, the the numerator above $E_t$ is just 1.
The $n$-point exchange integral can be recursively obtained by simply taking the residue of the above contact integral:
\eqs{
\mathcal{I}_I&\equiv\int_{-\infty}^{\infty} \frac{dp}{2\pi i} \frac{p^{-2}}{k_I^2+p^2}\Bar{\mathcal{C}}(k_1,\dots,ip)\Bar{\mathcal{C}}(ip,\dots,k_n),\\
\mathcal{I}_{IJ \dots M}&\equiv\int_{-\infty}^{\infty} \frac{dp}{2\pi i} \frac{p^{-2}}{k_I^2+p^2}\Bar{\mathcal{C}}(k_1,\dots,ip)\Bar{\mathcal{I}}_{J\dots M}(ip,\dots,k_n), \label{exchange}
}
where we define a shifted function: $\Bar{\mathcal{C}}=\mathcal{C}(k_1,\dots,ip)-\mathcal{C}(k_1,\dots,-ip)$.
We believe these cover all $n$-point scalar integrals for Gravity in $d=3$. We will provide more details and examples at five-point in Appendix \ref{appendixB}. Therefore, if one is provided with an $n$-point Mellin-Momentum amplitude, one can simply follow the map to obtain the wavefunction coefficients, requiring only the computation of a finite number of residues without the need for any time integrals. Then one can use the recent transitioning from AdS to dS\cite{Sleight:2021plv,Sleight:2020obc,DiPietro:2021sjt}, to obtain cosmological correlators.
 \section{Outlook}\label{section8}
In this letter, we show that the analytic structure of $n$-point Mellin-Momentum amplitudes is remarkably simple, and can be computed recursively like flat space amplitudes, which confirms the proposal that Mellin-Momentum amplitudes in (A)dS should play a similar role as S-matrix in flat space. Pragmatically, it is easy to check that our new result for five-graviton amplitude by bootstrap method is correct by construction, as any mistake would result in non-physical poles. This will give us the Gravity Quadrispectrum for cosmological correlators. We will present more details in a forthcoming paper\cite{JY2024} and an automated Mathematica file for the algorithm. The structure of five graviton also confirms that the double copy construction in \cite{Mei:2023jkb} can be extended to higher point, and it would be interesting to explore more about this in the future.

Given the simplicity and flat space structure of the $n$-point gluon amplitudes, our dream for the future is to discover an $n$-point formula akin to the Park-Taylor formula in the S-matrix\cite{Parke:1986gb}. To achieve this, we first need to formulate a spinor-helicity representation\cite{Maldacena:2011nz} in Mellin-Momentum space that makes the on-shell degrees of freedom manifest. We hope to report progress on this in the future.\\
It would also be interesting to extend the Gravity calculation to loop-level. Once employing our bootstrap approach to determine the Mellin-Momentum amplitude, we are then left with scalar loop integrals. Particularly, it was shown in \cite{Chowdhury:2023arc} that the scalar loop integral for the in-in correlator is closer to the S-matrix in flat space.\\
More generally, Our bootstrap approach does not rely on the spacetime symmetry\cite{Pajer:2020wxk} and the AdS study here is just the simplest example of curved space. Therefore, there is potential for its implementation in more general curved backgrounds, such as FLRW spacetime or even black hole backgrounds. We hope to explore this further in the future.\\
\begin{acknowledgments}
We would like to thank Chandramouli Chowdhury, Arthur Lipstein and Enrico Pajer for valuable comments on the draft. JM is supported by a Durham-CSC Scholarship. YM is supported by a Edinburgh Global Research Scholarship.
\end{acknowledgments}


\begin{widetext}
\appendix
\section{Flat space structure Feynman rules for Yang-Mills}\label{appendixA}
In this section, we will derive a new set of Feynman rules that makes the flat space structure manifest. We will employ the boundary transverse gauge, $k_{\mu} \cdot A^{\mu}=0$, which allows us to have only scalar-like propagators\cite{Armstrong:2022mfr}. By directly solving the equation of motion, we found the Feynman rules are identical to flat space in Coulomb gauge with the simple replacement of the EoM operator $\frac{1}{s}$ with $\frac{1}{\mathcal{D}^{d-1}_{k}}$:
\eqs{\begin{tikzpicture}
  \begin{feynman}
    \vertex (a) ;
    \vertex[right=1.5cm of a] (b);
    \node[above] at (a) {\(\mu\)};
    \node[above] at (b) {\(\nu\)};
    \diagram* {
      (a) -- [gluon] (b),
    };
  \end{feynman}
\end{tikzpicture}: G_{\mu \nu}&=\frac{\Pi_{\mu \nu}}{\mathcal{D}^{d-1}_{k}},  \\
\begin{tikzpicture}
  \begin{feynman}
    \vertex (a) ;
    \vertex[right=1.5cm of a] (b);
    \node[above] at (a) {\(z\)};
    \node[above] at (b) {\(z\)};
    \diagram* {
      (a) -- [gluon] (b),
    };
  \end{feynman}
\end{tikzpicture}: G_{zz} &=\frac{1}{z^2k^2},
}
where $\Pi_{\mu \nu}=\eta_{\mu \nu}-\frac{k_{\mu}k_{\nu}}{k^2}$ is the spin-1 projection tensor. Similarly, after employing the Mellin-Fourier transform $\phi(x,z)\sim e^{ik\cdot x}z^{-2s+d/2}\phi(k,s)$, it's easy to derive the color ordered vertex and see that they are the same as flat space with:
 \eqs{
 \begin{tikzpicture}[baseline=(current bounding box.center)]
  \begin{feynman}
    \vertex (a);
    \vertex [right=1cm of a] (b);
    \vertex [above right=1cm of b] (c);
    \vertex [below right=1cm of b] (d);
    \node[above left] at (a) {\(p_1\)};
    \node[below left] at (a) {\(m\)};
    \node[above right] at (c) {\(p_2\)};
    \node[below right] at (c) {\(n\)};
    \node[above right] at (d) {\(p_3\)};
    \node[below right] at (d) {\(q\)};
    \diagram* {
      (a) -- [gluon] (b),
      (b) -- [gluon] (c),
      (b) -- [gluon] (d),
    };
  \end{feynman}
\end{tikzpicture}:&  \;  V_{mnq}(p_1,p_2,p_3)
   =\frac{1}{2}\left(\eta _{mn}(p_1-p_2)_q +\eta _{nq}(p_2-p_3)_m+\eta _{qm}(p_3-p_1)_n\right),\\
   \begin{tikzpicture}[baseline=(current bounding box.center)]
  \begin{feynman}
    \vertex (a);
    \vertex [above left=1cm of a](b);
    \vertex [above right=1cm of a] (c);
    \vertex [below right=1cm of a] (d);
    \vertex [below left=1cm of a] (e);
    \node[above left] at (b) {\(m\)};
    \node[above right] at (c) {\(n\)};
    \node[below right] at (d) {\(q\)};
    \node[below left] at (e) {\(o\)};
    \diagram* {
      (a) -- [gluon] (b),
      (a) -- [gluon] (c),
      (a) -- [gluon] (d),
      (a) -- [gluon] (e),
    };
  \end{feynman}
\end{tikzpicture}: & \; V_{mnqo}  =\frac{1}{2} \eta _{mq}\eta _{no} -\frac{1}{4}(\eta _{mn}\eta _{qo}+\eta _{mo}\eta _{nq}) ,
}
where $p^m=(-2s+d/2,z k^{\mu})$. So we can easily uplift any Feynman diagrams from flat space to AdS amplitude. The crucial distinction between flat space and AdS is the $k_I$ pole, which is non-physical in flat space. However, this pole in AdS is precisely the signal of a CFT, which is necessary for the CFT to have an infinite expansion (corresponding to infinite number of descendent operators) in the OPE limit\cite{Arkani-Hamed:2015bza}. We have also used these Feynman rules to compare with the bootstrap result above, finding perfect agreement.
\section{Scalar Integrals for Gravity}\label{appendixB}
The scalar integrals for Yang-Mills in $d=3$ are simply plane waves, so we focus on Gravity here. Firstly, the inversion propagator in amplitude is defined by the insertion of the Green function as:
\begin{align}
    (\mathcal{D}(z))^{-1} \mathcal{O}(z)&= \int \frac{dy}{y^{d+1}} G(z,y) \mathcal{O}(y), \\
    \mathcal{D}_{k}^{\Delta}G(z,y)&=z^{d+1} \delta(z-y).
\end{align}
We will be using the following representation of Bulk-to-Bulk propagator \cite{Liu:1998ty}:
\eqs{
G(k,z_1,z_2)=\int_{0}^{\infty}\frac{dp}{2\pi i}\frac{-p^{d+1-2\Delta}}{k^2+p^2} \left( \phi_{\Delta}(z_1,i p)-\phi_{\Delta}(z_1,-i p)\right) \left( \phi_{\Delta}(z_2,i p)-\phi_{\Delta}(z_2,-i p)\right),
}
where $\phi_{\Delta}(z,k)=(z k)^{\Delta-d/2} K_{\Delta-d/2}(kz)$ is the usual Bulk to Boundary propagator. This makes the recursive relation \eqref{exchange} manifest. For example, the scalar integral with two propagators for 5-point graviton is:
    \eqs{
&\frac{1}{\mathcal{D}_{k_{12}}^d\mathcal{D}_{k_{45}}^d}\to\int_{-\infty}^{\infty}\frac{dp_1}{2\pi i} \frac{p_1^{-2}}{k_{12}^2+p_1^2}\Bar{\mathcal{C}}_1(k_1,k_2,ip_1)\Bar{\mathcal{I}}_{45}^{ (4)}(ip_1,k_3,k_4,k_5)\\
&=\underset{p_1}{\mathrm{Res}}\; \underset{p_2}{\mathrm{Res}}\; \frac{64 k_3^3 p_1^4 p_2^4 \left(k_1^2+4 k_2 k_1+k_2^2+p_1^2\right) \left(k_4^2+4 k_5 k_4+k_5^2+p_2^2\right)}{\pi ^2 \left(\left(k_1+k_2\right){}^2+p_1^2\right){}^2
   \left(k_{12}^2+p_1^2\right) \left(k_3^2+\left(p_1-p_2\right){}^2\right){}^2 \left(\left(k_4+k_5\right){}^2+p_2^2\right){}^2 \left(k_{45}^2+p_2^2\right)
   \left(k_3^2+\left(p_1+p_2\right){}^2\right){}^2}
}
where in the second step we can simply recycle the three-point contact and the four-point exchange results, and we are left with taking a few simple residues of $p_1,p_2$. This completes the mapping to momentum space without doing any integrals.
\end{widetext}

\bibliography{refs.bib} 

\end{document}